\documentclass[conference]{IEEEtran}
\IEEEoverridecommandlockouts

\usepackage{epsfig} 
\usepackage[cmex10]{amsmath}
\usepackage{amssymb}  
\usepackage[dvips]{color}
\usepackage[left=54pt,right=54pt,top=54pt,bottom=58pt]{geometry}

\newtheorem{theorem}{Theorem}

\newtheorem{assumption}{Assumption}
\graphicspath{{figs/}}

\newcommand{\dif}{\mathrm{d}}
\newcommand{\avg}{\mathrm{a}}

\begin{document}

\title{\vspace{10pt}Second-Order Newton-Based Extremum Seeking for Multivariable Static Maps}

\author{\IEEEauthorblockN{Azad Ghaffari}
\IEEEauthorblockA{\textit{Dept. of Mechanical Engineering} \\
\textit{Wayne State University}\\
Detroit, USA \\
aghaffari@wayne.edu}
\and
\IEEEauthorblockN{Tiago Roux Oliveira}
\IEEEauthorblockA{\textit{Dept. of Electronics and Telecommunication Engineering} \\
\textit{State University of Rio de Janeiro}\\
Rio de Janeiro, Brazil \\
tiagoroux@uerj.br}
}

\maketitle

\begin{abstract}

A second-order Newton-based extremum seeking (SONES) algorithm is presented to estimate directional inflection points for multivariable static maps. The design extends the first-order Newton-based extremum seeking algorithm that drives the system toward its peak point. This work provides perturbation matrices to estimate the second- and third-order derivatives necessary for implementation of the SONES. A set of conditions are provided for the probing frequencies that ensure accurate estimation of the derivatives. A  differential Riccati filter is used to calculate the inverse of the third-order derivative. The local stability of the new algorithm is proven for general multivariable static maps using averaging analysis. The proposed algorithm ensures uniform convergence toward directional inflection point without requiring information about the curvature of the map and its gradient. Simulation results show the effectiveness of the proposed algorithm.

\end{abstract}


\section{Introduction}

A century since its invention and first application \cite{Leblanc_1922}, and a quarter century since the proof of its convergence \cite{KW:2000}, extremum seeking (ES) is recognized as arguably the most valuable real-time optimization tool for model-free {\em dynamic} systems. Reference \cite{scheinker_100_years} highlights some of ES principal current advances in algorithm design, theory, and practical applications in several domains.

Confidence in ES, brought about by the proof of its stability for static and general nonlinear dynamic systems in continuous time \cite{KW:2000}, has engendered and accelerated the industrial uses of ES. Examples include chip manufacturing, wind and photovoltaic energy, mobile robots, bioengineering systems, acoustic systems, noncooperative games, neuromuscular electrical stimulation, biological reactors, oil-drilling systems, and flow-traffic control for urban mobility. These engineering applications have been further enriched by recent theoretical developments that have opened the door to the design and analysis of new types of ES, such as those modeled by discrete-time systems \cite{CKAL:2002}, multivariable systems \cite{GKN:2012}, non-cooperative games \cite{FKB:2012}, hybrid dynamical systems \cite{PT:2017}, stochastic systems \cite{MK:2009,LK:2010}, systems with delays \cite{OKT:2017,ZFO:2023}, and partial differential equations (PDEs) \cite{bib1}.

In particular, the authors of \cite{GKN:2012} present a multivariable Newton-based extremum seeking algorithm, which yields arbitrarily assignable convergence rates for each of the elements of the input vector. They generate the estimate of the Hessian matrix by generalizing the idea proposed in \cite{Nesic:2010} for the scalar case. After that, Mills and Krstic \cite{Greg:2018} generalize the earlier results to a scalar version of the Newton-based extremum seeking algorithm which maximizes the map's higher derivatives through measurements of the map. Hence, rather than optimizing the map itself, they offer a new scheme which can be useful in practice for maximizing the sensitivity of an input-to-output function to improve the signal-to-noise ratio as well as minimizing parameter sensitivity when designing feedforward controllers. For instance, the authors in \cite{Vinther:2013} present a refrigeration system where a suitable operating point is located at the maximum negative slope that is subject to change. This point of zero curvature corresponds to a minimum of the first derivative of the input-output map. Hence, being able to track the minimum of the first derivative in real-time would allow the system to operate almost the whole time at the most suitable operating condition. Following these ideas, Rusiti et al. \cite{EJC_2018} have expanded such domain application in order to include delays in the input and/or output channels of the maps to be optimized. 

This paper considers an open problem in the existing literature for multivariable ES. To the best of the author's knowledge, there has been no technical result which considers the real-time optimization though ES for the ``first-derivative'' of multi-input maps, or precisely describing, for finding the extremum of the gradient vector of a multivariable static function. By these reasons, the term {\it higher-order} ES might be suitable here. In this context, the proposed controller is called second-order ES since it differs from the ``first-order'' classical approach which optimizes only zero-derivatives (or the own function) for a given input-output map. This is a challenging problem since in the high-order multi-parametric search there can be zero columns of the Hessian matrix in the ``directional extremum'' (in every search direction, there may be local extremum points or even saddle points different from the global one). Therefore, along a directional search, isolated extremum points can exist. Zero columns of the Hessian may indicate directional inflection points, where the curvature of the map changes direction. In addition, this problem has interesting aspects, such as the conditions that ensure the uniqueness of the extremum point for gradients along all axes. Note that we are dealing with vector spaces with increasing dimensions as the order of derivative increases, which may end up creating combination of extremum and saddle points. This work investigates a new paradigm: the definition and estimation of extremum for non-scalar functions.

This work focuses on estimating the directional inflection point via a perturbation-based ES. The objective is finding the maximum of gradient along a priori known axis. In this event, if $y=h(\theta)$ with $\theta=[\theta_1~\theta_2~\cdots~\theta_p]^\top$ has an inflection point along $\theta_m$-axis, then the $m^\mathrm{th}$ column of the Hessian matrix is zero, $\partial^2 h(\theta)/\partial\theta\partial\theta_m=0$, and $\partial^3 h(\theta)/\partial^2\partial\theta_m<0$. A second-order gradient-based ES can be generated using the estimate of Hessian to derive the system toward the inflection point. Performance of the gradient-based algorithms is heavily impacted by the curvature of the function. Particularly, one needs information of the curvature rate to properly tune the gradient-based algorithm, which in the case of multivariable maps becomes an almost impossible task to perform. Therefore, this work proposes a second-order Newton-based ES (SONES) to ensure uniform transient is seeking directional inflection points.

The SONES requires an accurate estimate of the second-order derivative and the inverse of the third-order derivative of the multivariable map. The inverse of the third-order derivative is produced by feeding the initial estimate of the third-order derivative to a dynamic Riccati filter. One challenging aspect of this work is producing the perturbation matrix that generates an accurate estimate of the third-order derivative. For this purpose, the new choice of the probing frequencies are necessary. Unlike \cite{Greg:2018}, notice that such multiplicative dithers (or demodulation signals) are vectors rather than scalars for the multivariable case. The proposed perturbation matrix to estimate the Hessian of the map is a new variation of the matrix in \cite{GKN:2012}. A new perturbation matrix is presented to estimate the third-order derivative of the map. The number of conditions on the perturbation frequencies dramatically increases, but the proposed algorithm can successfully detect directional inflection points since the algorithm drives the system to the zero column of the Hessian. The proof utilizes the finite-horizon averaging theorem. Numerical simulations illustrate the obtained results.

Section~\ref{sec:derivatives} states the problem and presents the estimation of the second- and third-order derivatives. Section \ref{sec:SONES} presents the second-order Newton-based scheme and its stability analysis for the static map. Section \ref{sec:sim} presents an illustrative example to highlight the effectiveness of the SONES. Section~\ref{sec:con} concludes the paper. Appendix~\ref{apdx1} presents conditions on the probing frequencies. 

\section{High-Order Derivatives of Static Map}\label{sec:derivatives}

The conventional extremum seeking (ES) rely on driving the estimate of the gradient vector towards zero. Although the Newton-based ES removes the impact of the map curvature on the convergence rate of the algorithm, the gradient direction still determines system convergence towards the extremum point. Therefore, in this work, the conventional gradient- and Newton-based algorithms is referred to as {\bf first-order ES} algorithms, where the objective is to find the extremum value of an unknown map.

The objective of this work is formulating an extremum seeking algorithm to estimate the directional inflection point of a static map. In other words, it is desirable to find the maximum gradient point along a priori known axis.  Consider a differentiability static map of class $\mathcal{C}^r$ for $r\ge3$
\begin{equation}\label{eq:s-cost}
y=h(\theta),\quad \theta=\left[\begin{array}{cccc}\theta_1&\theta_2&\cdots&\theta_p\end{array}\right]^\top.\end{equation}
Denote $G_m(\theta)=\partial h(\theta)/\partial\theta_m$ for $m=1,2,\cdots,p$. Assume that $G_m(\theta)$ has a local maximum at $\theta^*$, which indicates that the curvature along $\theta_m$-axis changes direction at $\theta^*$. The inflection point corresponds to $\partial G_m(\theta)/\partial\theta=0$, i.e., $\partial^2 h(\theta)/\partial\theta\partial\theta_m=0$, which indicates that the $m^\mathrm{th}$ column of the Hessian matrix is zero at the inflection point.  The cost function is unknown in (\ref{eq:s-cost}), but $y$ can be measured and $\theta$ can be manipulated. Then, one can utilize prior work to estimate the Hessian and drive the system towards the inflection point. Therefore, the extremum seeking algorithm that determines the directional inflection point is called {\bf second-order ES}.  

Since the expansion of the gradient-based ES to estimate directional inflection point is straight forward, the details are left to the interested reader to explore. This work provides an algorithm and proof of stability of a second-order Newton-based ES for static maps. The proposed algorithm requires the estimates second- and third-order derivatives along $\theta_m$-axis.

\begin{figure}
\centering{
\includegraphics[width=\columnwidth, clip]{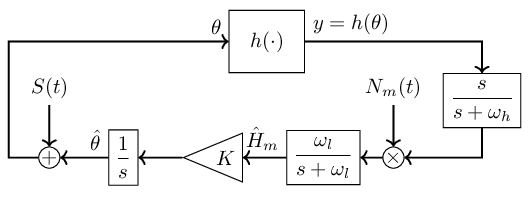}}
\vspace{-5mm}
\caption{Second-order gradient-based extremum seeking algorithm to find the directional inflection point along $\theta_m$-axis, where $K$ is a diagonal matrix with positive elements.}
\vspace{-5mm}
\label{fig:s-grad}
\end{figure}
The second-order gradient-based extremum seeking scheme that estimates the directional inflection point of a multivariable static map is shown in Fig.~\ref{fig:s-grad}, where $K$ is a positive diagonal matrix, and the perturbation input is given as
\setlength{\arraycolsep}{0.0em}
\begin{eqnarray}%
\label{eq:S}
S(t)&\;=\;&\left[\begin{array}{cccc}a_1\sin(\omega_1 t)&\quad \cdots&\quad a_n\sin(\omega_n t)\end{array}\right]^\top,
\end{eqnarray}%
where $\omega_i\in\mathbb{R}$ and $a_i$'s are small real numbers for $i\in\{1,2,\cdots,p\}$. The estimate of column $m$ of the Hessian matrix is the average of $N_m(t) y$. One can use equations (20) and (21) from \cite{GKN:2012} to generate $N_m(t)$. The probing frequencies need to satisfy
\vspace{-.1cm} \begin{equation}\omega_i\notin\Big\{\omega_j,\frac{1}{2}(\omega_j\!\!+\omega_k),\!\omega_j\!+\!2\omega_k, \omega_j\!+\!\omega_k\pm\omega_l\Big\},
\vspace{-.1cm}%
\end{equation}%
for all distinct $i,j,k,$ and $l$.  The low- and high-pass filter is added to improve the accuracy of the estimate of column $m$ of the Hessian matrix.

Although the prior work~\cite{GKN:2012} provides the perturbation matrix to estimate the Hessian of a map, this research revisits the Hessian perturbation matrix to establish similar patterns between perturbation matrices to generate estimates of the first-, second-, and third-order derivatives of the map.  Denote $\tilde\theta=\hat\theta-\theta^*$. Using $\theta=\hat\theta+S(t)$, one obtains $\theta=\theta^*+\tilde\theta+S(t)$. Using the Taylor series expansion of the cost function around the inflection point gives 
\setlength{\arraycolsep}{0.0em}
\begin{eqnarray}%
\label{eq:taylor}
y&=&h(\theta^*+\tilde{\theta}+S(t))\nonumber\\
~&=&h(\theta^*)+\sum_{i=1}^{p}\frac{\partial h(\theta^*)}{\partial\theta_i}\left(\tilde\theta_i+S_i\right)+\nonumber\\
~&+&\!\frac{1}{2}\sum_{i=1}^{p}\sum_{j=1}^p\frac{\partial^2 h(\theta^*)}{\partial\theta_i\partial\theta_j}\left(\tilde\theta_i+S_i\right)\!\!\left(\tilde\theta_j+S_j\right)+\nonumber\\
~&+&\!\frac{1}{6}\sum_{i=1}^{p}\sum_{j=1}^p\sum_{k=1}^{p}\frac{\partial^3 h(\theta^*)}{\partial\theta_i\partial\theta_j\partial\theta_k}\left(\tilde\theta_i\!+\!S_i\right)\!\!\left(\tilde\theta_j\!+\!S_j\right)\!\!\left(\tilde\theta_k\!+\!S_k\right)\!+\nonumber\\
~&+&R(\tilde{\theta}+S(t)),
\end{eqnarray}%
where $S_i=a_i\sin(\omega_it)$ for $i\in\{1,2,\cdots,p\}$ and $R(\tilde\theta+S(t))$ stands for higher order terms in $\tilde\theta+S(t)$. 

The average of the second-order derivative of~\eqref{eq:taylor} equals
\begin{eqnarray}
\frac{1}{\Pi}\int_0^\Pi \frac{\partial^2 y}{\partial\theta_i\partial\theta_j}\dif t&=&\frac{\partial^2 h(\theta^*)}{\partial\theta_i\partial\theta_j}+\sum_{k=1}^{p}\frac{\partial^3 h(\theta^*)}{\partial\theta_i\partial\theta_j\partial\theta_k}\tilde\theta_k+\nonumber\\
&+&\underbrace{\frac{1}{\Pi}\int_0^\Pi\frac{\partial^2 R(\tilde\theta+S(t))}{\partial\theta_i\partial\theta_j}\dif t}_{O(|\tilde\theta|^2,|a|^2)},
\end{eqnarray}
where $i, j=1,2,\cdots,p$ and $\Pi$ represents the averaging period. One can rewrite~\eqref{eq:taylor} as
\begin{eqnarray}
y&=&\frac{1}{2}\sum_{i=1}^p\sum_{j=1}^p\left(\frac{\partial h(\theta^*)}{\partial\theta_i\partial\theta_j}+\sum_{k=1}^p\frac{\partial^3 h(\theta^*)}{\partial\theta_i\partial\theta_j\partial\theta_k}\tilde\theta_k\right)S_iS_j+\nonumber\\
&+&F(S_1,\cdots,S_p,S_1^3,\cdots,S_lS_mS_n,\cdots,S_p^3)+\nonumber\\
&+&R(\tilde\theta+S(t)),
\end{eqnarray} 
where $F(\cdot)$ represents terms of the first- and third-order powers of the perturbation input $S(t)$. Note that
\begin{equation}
\sin(\omega_it)\sin(\omega_jt)=\frac{1}{2}\Big(\cos(\omega_i-\omega_j)t-\cos(\omega_i+\omega_j)t\Big).
\end{equation}
Then, one can generate an estimate of the second-order derivative by taking the average of $N(t) y$, where one option for the entries of $N(t)$ is as following
\begin{eqnarray}
N_{i,i}&=&-\frac{8}{a_i^2}\cos(2\omega_it)\\
N_{i,j}&=&-\frac{4}{a_ia_j}\cos(\omega_i+\omega_j)t, \quad i\neq j
\end{eqnarray}
for $i,j\i\{1,2,\cdots,p\}$. Proper choice of the probing frequencies will eliminate all the terms of $F(\cdot)N(t)$ in the average sense and gives 
\begin{eqnarray}
\label{eq:hessian}
\frac{1}{\Pi}\int_{0}^{\Pi}\!\!\!N_{i,j}(t)y\dif t=\frac{\partial^2 h(\theta^*)}{\partial\theta_i\partial\theta_j}+\sum_{k=1}^p\frac{\partial^3 h(\theta^*)}{\partial\theta_i\partial\theta_j\partial\theta_k}\tilde\theta_k+\nonumber\\
\underbrace{\frac{1}{\Pi}\int_{0}^{\Pi}\!\!\!R(\tilde\theta+S(t))N_{i,j}(t)\dif t}_{O(|\tilde\theta|^2,|a|^2)}.
\end{eqnarray}
Comparing the right-hand side of \eqref{eq:taylor} and \eqref{eq:hessian} shows that the average of $N(t)y$ provides an accurate estimate of the second-order derivative of $y=h(\theta)$ in the average sense.

Following similar approach, one can show that the average of the third-order derivative of the map gives
\begin{eqnarray}
\label{eq:thirdorder}
\frac{1}{\Pi}\int_0^\Pi \frac{\partial^3 y}{\partial\theta_i\partial\theta_j\partial\theta_k}\dif t &=&\frac{\partial^3 h(\theta^*)}{\partial\theta_i\partial\theta_j\partial\theta_k}+\nonumber\\
&+&\underbrace{\frac{1}{\Pi}\int_0^\Pi \frac{\partial^3 R(\tilde\theta+S(t))}{\partial\theta_i\partial\theta_j\partial\theta_k}}_{O(|\tilde\theta|,|a|)}.
\end{eqnarray}
Moreover, \eqref{eq:taylor} can be rewritten as
\begin{eqnarray}
y&=&\frac{1}{6}\sum_{i=1}^p\sum_{j=1}^p\sum_{k=1}^p\frac{\partial^3 h(\theta^*)}{\partial\theta_i\partial\theta_j\partial\theta_k}S_iS_jS_k+\nonumber\\
&+&F'(S_1,\cdots,S_p,S_1^2,\cdots,S_p^2)+R(\tilde\theta+S(t)).
\end{eqnarray}
The following trigonometric identity holds
\begin{eqnarray}
&&\sin(\omega_it)\sin(\omega_jt)\sin(\omega_kt)=\frac{1}{4}\Big(\sin(\omega_i-\omega_j+\omega_k)t+\nonumber\\
&&+\sin(\omega_i+\omega_j-\omega_k)t-\sin(\omega_i-\omega_j-\omega_k)t-\nonumber\\
&&-\sin(\omega_i+\omega_j+\omega_k)t\Big).
\end{eqnarray}
An estimate of the third-order derivative of the map can be generated by taking the average of $P(t)y$, where one option for the entries of $P(t)$ is obtained as
\begin{eqnarray}
\label{eq:Piii}
P_{i,i,i}&=&-\frac{48}{a_i^3}\sin(3\omega_it)\\
\label{eq:Piij}
P_{i,i,j}&=&-\frac{16}{a_i^2a_j}\sin(2\omega_i+\omega_j)t, \quad i\neq j\\
\label{eq:Pijk}
P_{i,j,k}&=&-\frac{8}{a_ia_ja_k}\sin(\omega_i+\omega_j+\omega_k)t, \quad i\neq j\neq k,
\end{eqnarray}
for $i, j, k\in\{1,2,\cdots,p\}$. An appropriate choice of the probing frequencies will eliminate $F'(\cdot)P(t)$ in the average sense and provides
\begin{eqnarray}
\label{eq:Py}
\frac{1}{\Pi}\int_0^\Pi P_{i,j,k}(t)y\dif t&=&\frac{\partial^3 h(\theta^*)}{\partial\theta_i\partial\theta_j\partial\theta_k}+\nonumber\\
&+&\underbrace{\frac{1}{\Pi}\int_0^\Pi P_{i,j,k}(t)R(\tilde\theta+S(t))\dif t}_{O(|\tilde\theta|,|a|)}.
\end{eqnarray}
Comparing the righ-hand side of \eqref{eq:thirdorder} and \eqref{eq:Py} shows that the average of $P(t)y$ provides an accurate estimate of the third-order derivative of $y=h(\theta)$.
The averaging period is obtained from
\begin{equation}%
\label{eq:Pi}
{\Pi}=2\pi\times\mathrm{LCM}\Big\{\frac{1}{\omega_i}\Big\},\quad i\in\{1, 2, \cdots, p\},
\end{equation}%
where LCM stands for the least common multiple.

\section{Second-Order Newton-Based ES}\label{sec:SONES}

Consider a multivariable static map
\begin{eqnarray}
\label{eq:map}
y=h(\theta), \quad y\in\mathbb{R}, \theta\in\mathbb{R}^p.
\end{eqnarray}
The objective is to develop a feedback mechanism which determines the directional inflection point of $y$ without requiring the knowledge of either the inflection point, $\theta^*$, or the functions $h(\dot)$. 
The gradient vector of $y=h(\theta)$ is defined as
\begin{eqnarray}
G(\theta) = [G_1(\theta)~G_2(\theta)~\cdots~G_p(\theta)]^\top,
\end{eqnarray}
where  $G_m(\theta)=\partial h(\theta)/\partial\theta_m, m=1,2,\cdots,p$. Denote $H=\partial G(\theta)/\partial\theta=\partial^2 h(\theta)/\partial\theta^2$. Column $m$ of matrix $H$ is the gradient of $G_m(\theta)$.
\begin{assumption}
\label{assumption1}
There exists a inflection point $\theta^*$ along $\theta_m$-axis that satisfies
\begin{eqnarray}
\frac{\partial}{\partial\theta} G(\theta)&=&H, \quad H^\top=H\\
\frac{\partial}{\partial\theta} G_m(\theta^*)&=&H_m=0\\
\frac{\partial^2 }{\partial\theta^2} G_m(\theta^*)&=&T_m<0, \quad T_m^\top=T_m
\end{eqnarray}
\end{assumption}

Assumption~\ref{assumption1} suggests that $H_m$ vanishes at the inflection point. Hence, the estimate of $H_m$ can be used to drive the system towards the inflection point using the gradient-based parameter update law shown in Fig.~\ref{fig:s-grad}. The perturbation vector estimating $H_m$ is given as
\begin{eqnarray}
\label{eq:Nm}
N_m(t)&=&-\frac{4}{a_m}\Big[\frac{\cos(\omega_1+\omega_m)t}{a_1} \cdots \frac{2\cos(2\omega_m)t}{a_m} \cdots \nonumber\\
&&\hspace{1cm}\frac{\cos(\omega_p+\omega_m)t}{a_p}\Big]^\top.
\end{eqnarray}
The proof of stability of the second-order gradient-based algorithm is a straight forward expansion of the conventional (first-order) gradient-based ES.

\begin{figure}
\includegraphics[width=\columnwidth, clip]{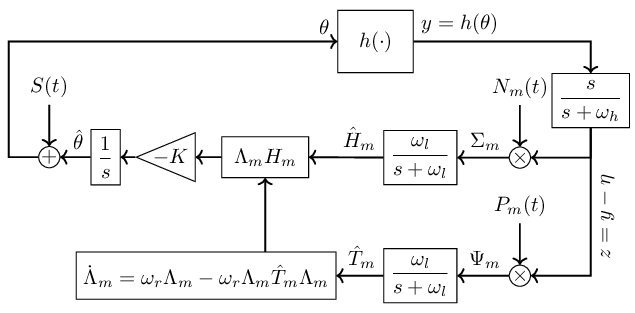}
\vspace{-5mm}
\caption{Second-order Newton-based ES proposed for estimating the inflection point along $\theta_m$-axis.}
\label{fig:s-newton}
\vspace{-5mm}
\end{figure}
Since Newton-based adaptive law provides uniform transient and reduces control calibration efforts, this work proposes a second-order Newton-based ES for estimating directional inflection points. A block diagram of the algorithm is shown in Fig.~\ref{fig:s-newton}. The perturbation vector $N_m(t)$ is given by \eqref{eq:Nm} and perturbation matrix $P_m(t)$ is  given as
\setlength{\arraycolsep}{2pt}
\begin{eqnarray}
\label{eq:Pm}
P_m(t)=
\begin{bmatrix}
P_{m,1,1} & P_{m,1,2} 	& \cdots & P_{m,1,p}\\
P_{m,2,1} & P_{m,2,2} 	& \cdots & P_{m,2,p}\\
\vdots 	& \vdots 		& \ddots & \vdots\\
P_{m,p,1} & P_{m,p,2} 	& \cdots & P_{m,p,p}
\end{bmatrix},
\end{eqnarray}
where the entries are obtained from~\eqref{eq:Piii}--\eqref{eq:Pijk}.

The probing frequencies $\omega_i$'s, the filter coefficients $\omega_h,$  $\omega_l,$ and $\omega_r$ and gain $K$ are selected as
\begin{eqnarray}
\omega_h &=&\delta\omega_h' = O(\delta)\\
\omega_l &=&\delta\omega_l' = O(\delta)\\
\omega_r &=&\delta\omega'_r = O(\delta)\\
K &=& \delta K''=O(\delta),
\end{eqnarray}
where $\delta$ is a small positive constant, $\omega_i$ is a real number, $\omega_h',$ $\omega_l',$ and $\omega_r'$ are $O(1)$ positive constants, and $K'$ is a $p\times p$ diagonal matrix with $O(1)$ positive elements.  ignoring these conditions shifts the estimate of the parameter away from its true value, and leading to inaccurate estimates of the second-  and third-order derivatives.

The system in Fig. \ref{fig:s-newton} is summarized as
\setlength{\arraycolsep}{2pt}
\begin{eqnarray}%
\label{eq:sysaa}
&&\frac{\dif}{\dif t}\begin{bmatrix}
\tilde{\theta}^\top & \hat{H}_m^\top & \tilde{\Lambda}_m^\top & \tilde{T}_m^\top & \tilde{\eta}
\end{bmatrix}^\top=\nonumber\\
&&
\left[\!\begin{array}{c}
-K(\tilde{\Lambda}_m+T_m^{-1})\hat{H}_m\\
-\omega_l\hat{H}_m+\omega_l\Big(y-h(\theta^*)-\tilde{\eta}\Big)N_m(t)\\
\omega_r\left(\tilde{\Lambda}_m+T_m^{-1}\right)\!\!\Big(I\! -\! (\tilde{T}_m+T_m)(\tilde{\Lambda}_m+H_m^{-1})\Big)\\
-\omega_l\left(\tilde{T}_m+T_m\right)+\omega_l\Big(y-h(\theta^*)-\tilde{\eta}\Big)P_m(t)\\
-\omega_h\tilde{\eta}+\omega_h\Big(y-h(\theta^*)\Big)
\end{array}\!\right]\!\!,
\end{eqnarray}
where error variables are introduced as
$\tilde{\theta}=\hat{\theta}-\theta^*,~ \tilde{\eta}=\eta-h(\theta^*),~\tilde{\Lambda}_m=\Lambda_m-\!T_m^{-1}$, and $\tilde{T}_m=\hat{T}_m-T_m$, and $\theta=\hat{\theta}+S(t)$.
A slight abuse of notation is performed by stacking matrix quantities $\tilde{\Lambda}_m$ and $\tilde{T}_m$ along with vector quantities, as alternative notational choices would be more cumbersome.

The average system is obtained as
\setlength{\arraycolsep}{0.0em}\begin{eqnarray}
\label{eq:sysavg}
&&\hspace{-.5cm}\frac{\dif}{\dif t}\left[\begin{array}{ccccc}\tilde{\theta}^{\avg \top}&\quad\hat{H}_m^{\avg \top}&\quad\tilde{\Lambda}_m^{\avg \top}&\quad\tilde{T}_m^{\avg \top}&\quad\tilde{\eta}^\avg\end{array}\right]^\top
=\nonumber\\
&&\hspace{-.5cm}
\delta\left[\begin{array}{c}
-K'(\tilde{\Lambda}_m^\avg+T_m^{-1})\hat{H}_m^\avg\\
-\omega_l'\hat{H}_m^\avg+\omega_l'\frac{1}{\Pi}\int_0^\Pi\nu(\tilde\theta^\avg+S(t))N_m(t)\dif t\\
\omega_r'(\tilde{\Lambda}_m^\avg+T_m^{-1})\big(I+(\tilde{T}_m^\avg+T_m)(\tilde{\Lambda}_m^\avg+T_m^{-1})\big)\\
-\omega_l'\big(\tilde{T}_m^\avg+T_m)+\omega_l''\frac{1}{\Pi}\int_0^\Pi\nu(\tilde\theta^\avg+S(t))P_m(t)\dif t\\
-\omega_h'\tilde{\eta}^\avg+\omega_h'\frac{1}{\Pi}\int_0^\Pi\nu(\tilde{\theta}^\avg+S(t))\dif t
\end{array}\right],
\end{eqnarray}%
where $\nu(q)=h(\theta^*+q)-h(\theta^*)$.
In view of Assumption~\ref{assumption1}, $\partial \nu(0)$${\partial^2\nu(0)}/{\partial q\partial q_m}=0$, ${\partial^3\nu(0)}/{\partial q^2\partial q_m}=T_m<0$. The stability of the average system is summarized in the following theorem.

\begin{theorem}\label{thm:avr}
Consider system~\eqref{eq:sysavg} under Assumption \ref{assumption1}. There exist $\bar{\delta}, \bar{a}>0$ such that for all $\delta\in(0,\bar{\delta})$ and $|a|\in(0,\bar{a})$ system~\eqref{eq:sysavg} has a unique exponentially stable periodic solution $\big(\tilde{\theta}^{{\Pi}}(t),\hat{H}_m^{{\Pi}}(t),\tilde{\Lambda}_m^{{\Pi}}(t),\tilde{T}_m^{{\Pi}}(t),\tilde{\eta}^{{\Pi}}(t)\big)$ of period ${\Pi}$ and this solution satisfies
\begin{eqnarray}
&&\Big|\tilde{\theta}_{i}^{{\Pi}}(t)-\sum_{j=1}^{p}c_j^ia_j^2\Big|\le O(\delta+\left|a\right|^3),\\ 
&&\Big|\hat{H}_m^{{\Pi}}(t)\Big|\le O(\delta),\\ 
&&\Big|\tilde{T}_m^{{\Pi}}(t)\Big|\le O(\delta+\left|a\right|^2),\\ 
&&\Big|\tilde{\eta}^{{\Pi}}(\tau)-\sum_{i=1}^p\frac{\partial \nu(0)}{\partial q_i}\sum_{j=1}^pc_j^ia_j^2-\nonumber,\\
&&\qquad\qquad-\frac{1}{4}\sum_{i=1}^{p}\frac{\partial^2\nu(0)}{\partial q_i^2}a_i^2\Big|\le O(\delta+\left|a\right|^4),\\
&&\big(\tilde{T}_m^\Pi(t)+T_m\big)\big(\tilde\Lambda_m^\Pi(t)+T_m^{-1}\big)=I
\end{eqnarray}
for all $t\ge0$, where
\setlength{\arraycolsep}{2pt}
\begin{eqnarray}%
\left[\begin{array}{cccc}c_{j}^1 & c_{j}^{2}& \cdots &c_{j}^{p}\end{array}\right]^\top=-\frac{1}{2}T_m^{-1}\frac{\partial^4}{\partial q\partial q_j^2\partial q_m}\nu(0)
\end{eqnarray}
for all $i,j\in\!\!\{1, 2, \cdots, p\}$.
\end{theorem}

Proof of Theorem~\ref{thm:avr} is omitted due to space constraints. The following provides a sketch for the proof. The Jacobian matrix of the averaged system~\eqref{eq:sysavg} at the equilibrium point is obtained as
\begin{eqnarray}
J^{\mathrm{a,e}}=\delta\begin{bmatrix} A_{1,1} ~&~ 0_{2p\times (2p+1)}\\ A_{2,1} ~&~ A_{2,2}
\end{bmatrix},
\end{eqnarray}
where $A_{22}$ is Hurwitz and one can show that
\begin{eqnarray}
A_{1,1}=\begin{bmatrix}0_{p\times p} ~&~ -K'\big(\tilde\Lambda_m^\avg+T_m^{-1}\\ \frac{\omega_l'}{\Pi}\int_0^\Pi \frac{\partial \nu}{\partial q}N_m\dif t ~&~ -\omega_l'I_{p\times p}\end{bmatrix}
\end{eqnarray}
is Hurwitz for small $|a|$. Since $J^\mathrm{a,e}$ is lower-triangular and $A_{1,1}$ and $A_{1,2}$ are Hurwitz, then $J^\mathrm{a,e}$ is Hurwitz and the equilibrium of the average system~\eqref{eq:sysavg} is exponentially stable if all elements of vector $a$ are sufficiently small. The proof can be completed using the averaging theorem~\cite{Khalil:1996}.

\section{Simulation Results}\label{sec:sim}

To illustrate the results and highlight the effectiveness of the second-order Newton-based extremum seeking method, the following static input-output map is considered 
\begin{eqnarray}
y&=&h(\theta)\nonumber\\
&=&1+(\theta_1-\theta_1^*)-1*(\theta_2-\theta_2^*)+\frac{3}{2}(\theta_2-\theta_2^*)^2-\nonumber\\
&-&\frac{1}{6}\Big(2(\theta_1-\theta_1^*)^3+3(\theta_1-\theta_1^*)^2(\theta_2-\theta_2^*)+\nonumber\\
&+&12(\theta_1-\theta_1^*)(\theta_2-\theta_2^*)^2+(\theta_2-\theta_2^*)^3\Big),
\end{eqnarray}
where $\theta^*=[1~~2]^\top$. The map has a inflection point along $\theta_1$-axis at $\theta^*$, i.e., $G_1(\theta)=\partial h(\theta)/\partial\theta_1$ has a maximum at $\theta^*$, which in turn leads to $H_1=\partial^2 h(\theta^*)/\partial\theta\partial\theta_1=0$ and 
\begin{eqnarray}
T_1=\frac{\partial^2}{\partial^2\theta\partial\theta_1}G_1(\theta^*)&=&\begin{bmatrix}-2~&~-1\\-1~&~-4\end{bmatrix}<0\\
\Lambda_1=T_1^{-1}&=&\begin{bmatrix}-0.57~~ & 0.14\\ 0.14~~ & -0.29\end{bmatrix}.
\end{eqnarray}

The test is performed with the following parameters, $\omega=[500~~300]^\top$~rad/s, $\omega_h=\omega_l=\omega_r=1$~rad/s, $a=[0.1~~0.1]^\top$, $K=\mathrm{diag}([0.02~~0.02])$, $T_1(0)=\mathrm{diag}([-50~~-50])$, $\Lambda_1(0)=T_1(0)^{-1}$, $\theta_0=[0~~0]^\top$. Note that $T_1(0)$ is a diagonal matrix with entries of large absolute values to avoid numerical singularities during the transient.

\begin{figure}[!t]
\centering
{\bf (a)} \hspace{1.5in} {\bf (b)}
\includegraphics[width=0.5\columnwidth]{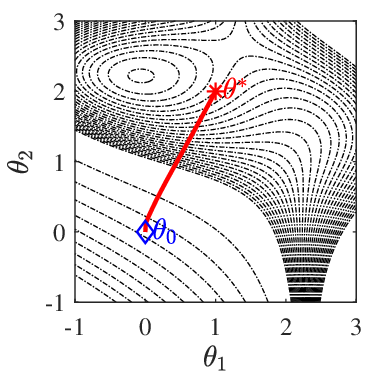}\includegraphics[width=0.5\columnwidth]{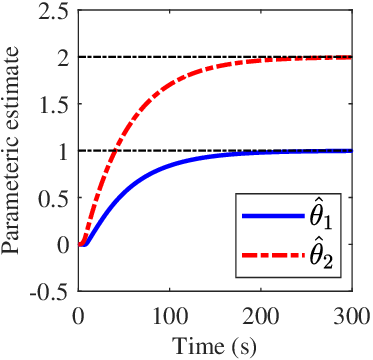}
\caption{ (a) The second-order Newton-based ES algorithm governs the system trajectory toward the inflection point along $\theta_1$-axis at $\theta^*=[1~~2]^\top$. The level sets of $y=h(\theta)$ indicate that the inflection point is near a local minimumu point and a saddle point at $[-0.07~~2.22]^\top$ and $[1.26~~ 2.62]^\top$, respectively. (b) The estimate of the parameters versus time.}
\label{fig:cost}
\end{figure}

\begin{figure}[!t]
\centering
{\bf (a)} \hspace{1.5in} {\bf (b)}
\includegraphics[width=.5\columnwidth]{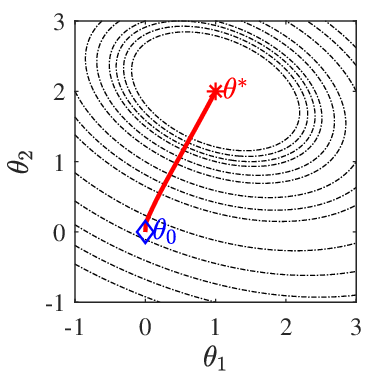}\includegraphics[width=.5\columnwidth]{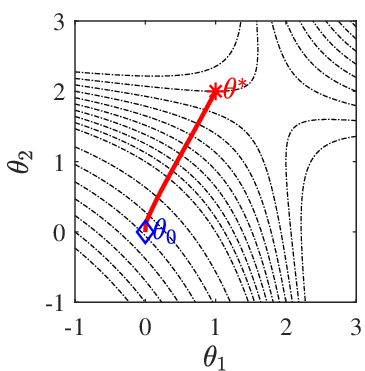}
\caption{(a) System trajectory among level sets of $G_1(\theta)=\partial h(\theta)/\partial\theta_1$. The inflection point along $\theta_1$-axis leads to a maximum point at $\theta^*$. (b) System trajectory among level sets of $G_2=\partial h(\theta)/\partial\theta_2$ indicating no extremum point.}
\label{fig:Glevels}
\end{figure}

\begin{figure}[!t]
\centering
\includegraphics[width=\columnwidth]{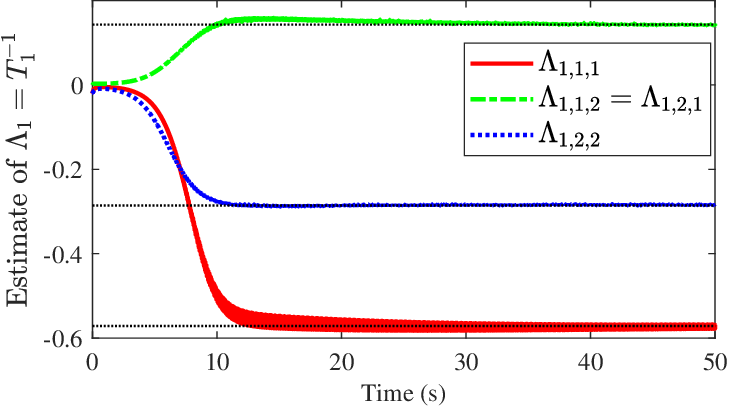}
\caption{Time evolution of the estimation of $\Lambda_1=T_1^{-1}$. The true value of $\Lambda_1$ is reached under $15$ seconds.}
\label{fig:Lambda}
\end{figure}

The map has an inflection point at $\theta^*$ along $\theta_1$-axis near a local minimum and saddle point that are respectively located at $[-0.07~~2.22]^\top$ and $[1.26~~ 2.62]^\top$. Fig. \ref{fig:cost}(a) illustrates the system trajectory among the level sets of the map. The algorithm successfully drives the system toward the inflection point. Evolution of the parameters versus time is depicted in Fig. \ref{fig:cost}(b). Fig.~\ref{fig:Glevels}(a) shows that the algorithm maximizes the gradient along $\theta_1$-axis. Fig.~\ref{fig:Glevels}(b) shows that the gradient along $\theta_2$-axis has a saddle point at $[1.8~~1.8]^\top$. Fig.~\ref{fig:Lambda} verifies that $\Lambda_1=T_1^{-1}$ converges to its actual value under $15$ seconds. 

\section{Conclusions}\label{sec:con}

Estimating directional inflection point of multivariable static maps can be achieved using a second-order gradient-based extremum seeking that uses the estimate of directional second-order derivative instead of gradient vector. However, the gradient-based estimation algorithms generally lead to low transient performance since their convergence rate is heavily impacted by the function curvature. The problem becomes exacerbated when the objective is estimating the directional inflection point since the curvature of the gradient determines convergence rate. As the number of parameters grow, achieving a uniform convergence rate of all parameters will become a cumbersome task. The Newton-based algorithm, which relies on the estimation of two consecutive derivatives at the same time, removes the trial and error process to update all parameters uniformly. A proper choice of probing frequencies along with careful selection of corner frequency of the high-pass, low-pass, and Riccati filters ensures accurate estimation of the second-order derivative and the inverse of the third-order derivative. The future work will expand the algorithm to dynamic maps.

\appendices
\section{Conditions on Probing Frequencies}
\label{apdx1}

The following conditions must be met for the probing frequencies to obtain an accurate estimate of the second- and third-order derivatives of the map.
\setlength{\arraycolsep}{3pt}
\begin{eqnarray}
\omega_i&\in&\Omega ~\mathrm{for}~i=1,2,\cdots,p\\
\Omega&=&\big\{\omega_i\in\mathbb{R}_{>0} ~|~ \nonumber\\
&&\omega_i\neq\omega_j \land \omega_i\neq 2\omega_j \land\omega_i\neq 3\omega_j \land \omega_i\neq 5\omega_j \land \nonumber\\
&& \omega_i\neq\omega_j+\omega_k \land \omega_i\neq\omega_j+\omega_k+\omega_l \land \nonumber\\
&&\omega_i\neq\omega_j+2\omega_k \land \omega_i\neq\omega_j+4\omega_k \land \nonumber\\
&&\omega_i\neq2\omega_j+3\omega_k \land \nonumber\\
&&\omega_i\neq\omega_j+2\omega_k+2\omega_l \land \omega_i\neq\omega_j+\omega_k+3\omega_l\nonumber
\end{eqnarray}
\begin{eqnarray}
&&\omega_i\neq\omega_j+\omega_k+\omega_l+2\omega_m \land \nonumber\\
&&\omega_i\neq\omega_j+\omega_k+\omega_l+\omega_m+\omega_n \land \nonumber\\
&&\omega_i\neq\frac{\omega_j+\omega_k}{2} \land \omega_i\neq\frac{\omega_j+3\omega_k}{2} \land \nonumber\\
&&\omega_i\neq\frac{\omega_j+2\omega_k}{3} \land \omega_i\neq\frac{\omega_j+\omega_k+\omega_l}{3} \land\nonumber\\
&&\omega_i\neq\frac{\omega_j+\omega_k}{4} \land \omega_i\neq\frac{\omega_j+\omega_k+2\omega_l}{2} \land \nonumber\\
&&\omega_i\neq\frac{\omega_j+\omega_k+\omega_l+\omega_m}{2} \land \nonumber\\
&&\omega_i+\omega_j\neq\omega_k+\omega_l \land \omega_i+\omega_j\neq\omega_k+3\omega_l \land \nonumber\\
&&\omega_i+\omega_j\neq2\omega_k+2\omega_l \land \omega_i+\omega_j\neq\omega_k+\omega_l+2\omega_m \land \nonumber\\
&&\omega_i+\omega_j\neq\omega_k+\omega_l+\omega_m+\omega_n \land \nonumber\\
&&\omega_i+2\omega_j\neq\omega_k+2\omega_l \land \omega_i+2\omega_j\neq \omega_k+\omega_l+\omega_m \land \nonumber \\
&&\omega_i+\omega_j+\omega_k\neq\omega_l+\omega_m+\omega_n\nonumber\\
&&\forall i\neq j\neq k\neq l\neq m\neq n\in\left\{1,2,\cdots,p\right\}\big\}
\end{eqnarray}

\bibliographystyle{IEEEtranS}

\end{document}